\documentclass[article,fleqn,usenatbib,useAMS]{ray}%

\usepackage{graphicx}
\usepackage{amsmath}	
\usepackage{amssymb}	
\usepackage{multicol}       
\usepackage{bm}		
\usepackage{pdflscape}	
\usepackage{aas_macros}
\usepackage{hyperref} 
\usepackage[T1]{fontenc}
\usepackage{ae,aecompl}

\usepackage{xcolor}

\newcommand{\lsun}{\ifmmode{{\rm ~L}_\odot}\else{~L$_\odot$}\fi}

\newcommand{\degr}{$^{\circ}$}

\newcommand{\ujybm}{\,$\mu$Jy/beam}

\def\apjl{{ApJ}}                
\def\apjs{{ApJS}}               
\def\ao{{Appl.~Opt.}}           
\def\aap{{A\&A}}                

\def\mnras{{MNRAS}}             
\def\pasa{{PASA}}               
\def\pasp{{PASP}}               
\title[Odd Radio Circles and their Environment]{Odd Radio Circles and their Environment}

\author[Norris, Crawford, \& Macgregor]{Ray P. Norris $^{1,3*}$, Evan Crawford $^{2}$ and Peter Macgregor $^{1,3}$
\affil{$^{1}$Western Sydney University, Locked Bag 1797, Penrith, NSW 2751, Australia}
\affil{$^{2}$School of Computer, Data and Mathematical Sciences, Western Sydney University, Locked Bag 1797, Penrith, NSW 2751, Australia}
\affil{$^3$CSIRO Space \& Astronomy, P.O. Box 76, Epping, NSW 1710, Australia}
}

\hypersetup{colorlinks,citecolor=blue,linkcolor=blue,urlcolor=blue}

\begin{document}
\begin{frontmatter}
\maketitle
\begin{abstract}
Odd Radio Circles (ORCs) are unexpected faint  circles of diffuse radio emission discovered in recent wide deep radio surveys. They are typically about one arcmin in diameter, and may be spherical shells of synchrotron emission about a million light years in diameter, surrounding galaxies at a redshift of $\sim0.2-0.6$. 
Here we study the properties and environment of the known ORCs.
All three known single ORCs either lie in a significant overdensity or have a close companion. 
If the ORC is caused by an event in the host galaxy, then the fact that they tend to be in an overdensity, or have a close companion,  may indicate that the environment is important in creating the ORC phenomenon, possibly because of an increased ambient density or magnetic field. 
\end{abstract}

\end{frontmatter}

\section{Introduction}

\begin{figure*}
\begin{center}
\includegraphics[height=9cm]{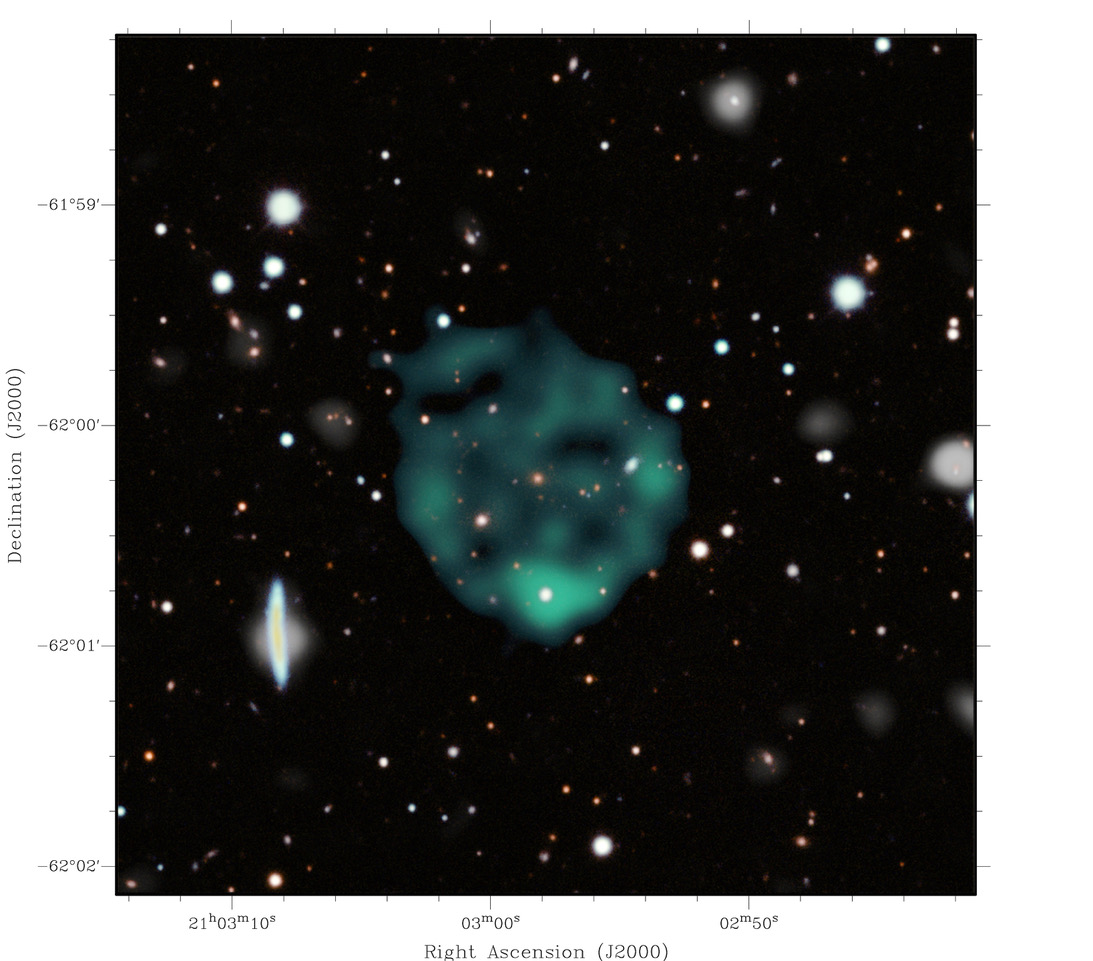}
\caption{
The ASKAP image of ORC1, adapted from \citet{norris20}. The resolution is 11 arcsec, and the rms sensitivity is 25 \ujybm, enhanced to show faint features, particularly the internal structure or ``spokes'' of the ORC.Radio data is shown in green, and the background optical image is taken from the Dark Energy Survey DR1 \citep{abbott18}. For the latter, the DES bands g, r, i, and z were assigned turquoise, magenta, yellow and red, respectively and combined using GIMP.  The optical/NIR image and the radio image were then combined using a masking technique. Image credit: Jayanne English.}
\label{askap1}
\end{center}
\end{figure*}

Odd Radio Circles (ORCs) are recently discovered circles of diffuse steep-spectrum radio emission with no corresponding diffuse emission at optical, infrared, or X-ray wavelengths. The first three ORCs were  found \citep{norris20} in the  Pilot Survey \citep{norris21} of the Evolutionary Map of the Universe \citep{emu}, using the Australian Square Kilometre Array Pathfinder (ASKAP) \citep{hotan21}. A fourth was found \citep{norris20} in data from the GMRT \citep{gmrt} and a fifth was found, also in ASKAP data, by \citet{koribalski21}. Figure \ref{askap1} shows an ASKAP image of the first ORC to be discovered.

The ORCs superficially resemble supernova remnants (SNRs), but their  Galactic latitude distribution is inconsistent with that of SNRs \citep{norris20}. Similarly, other explanations involving known classes of objects (e.g. starburst rings, gravitational lenses, etc.) have also been shown to be inconsistent with the data \citep{norris20}. Instead, we are forced to search for other explanations for this phenomenon. 

\begin{figure*}
\includegraphics[height=6cm]{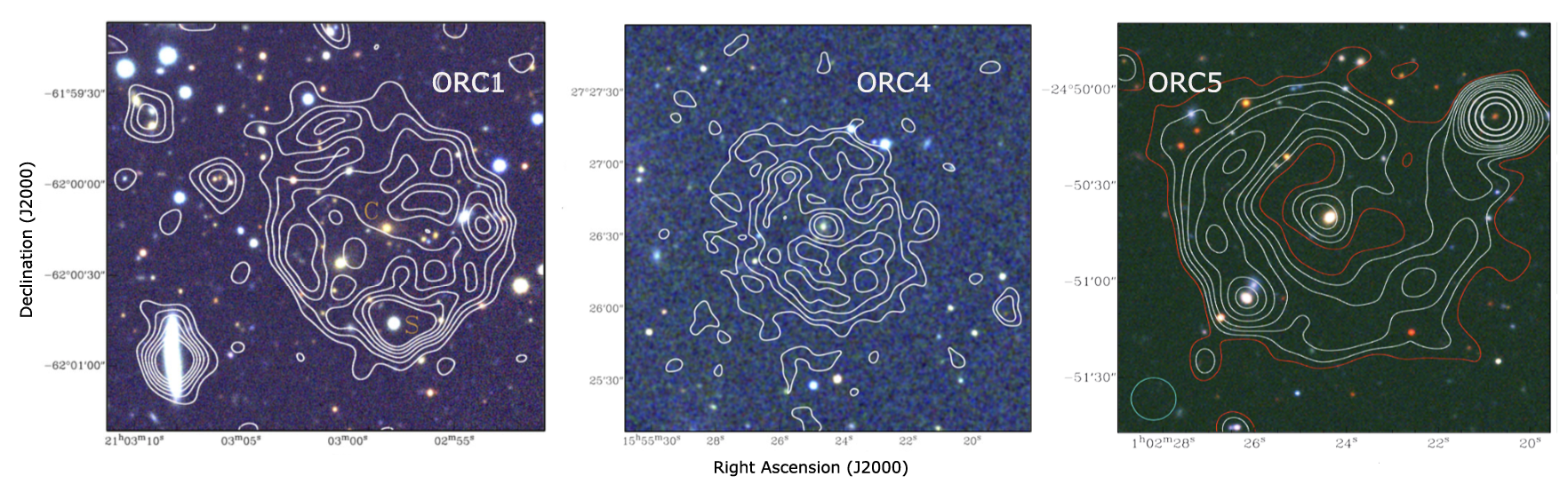}
\caption{(Left) 
The ASKAP radio image of ORC1,
adapted from \citet{norris20}. The contours show the radio data (at 45, 90, 135, 180, 225, and 270 \ujybm) overlaid onto a DES \citep{abbott18} 3-color composite image; DES gri-bands are colored blue, green, and red, respectively. 
(Centre) 
The GMRT radio  image of ORC 4, adapted from \citet{norris20}. The contours show the radio data (at  150, 250, 400, 600, and 800 \ujybm) overlaid onto a SDSS 3-color composite image; SDSS gri-bands are colored blue, green, and red, respectively.
(Right) 
The ASKAP radio image of ORC5,
adapted from \citet{koribalski21}. The  contours show the radio data overlaid onto a DES 3-color (grz) composite image. The dark red contour is at 65 \ujybm\ and remaining contours are at  90, 120, 170, 220, 270, 400, 600, and 800 \ujybm. The green ellipse shows the size of  the synthesised beam.}
\label{3orcs}
\end{figure*}

Of the five known ORCs, three are single, and share similar parameters, and are shown in Figure \ref{3orcs}. 
At the centre of each of the three single ORCs, there is a galaxy visible in optical wavelengths (see Figure \ref{3orcs}).  The probability of this happening by chance  is so low as to be extremely unlikely \citep{norris22}. We therefore call this galaxy the ``host galaxy'' of the ORC. The properties of the three single ORCs are summarised in Table \ref{table}.

\begin{table*}
	\centering
	\caption{Properties of the three single ORCs }
	\label{table}
	\begin{tabular}{cccccccl} 
		\hline
		Short & IAU &  RA (deg) & Dec (deg) & host & redshift   \\
		 name  &  name    &  J2000    &  J2000  & galaxy  &     \\
		\hline
		ORC~1 & ORC J2103-6200 & 315.74292 & --62.00444 & WISEA J210258.15–620014.4 & 0.551   \\
        ORC~4 & ORC J1555+2726 & 238.85272 &+27.44271 & WISEA J210258.15–620014.4 &  0.457     \\
        ORC~5 & ORC J0102-2450 & 15.60208 &-24.84392 & WISEA J010224.35–245039.6 & 0.270 \\
		\hline
	\end{tabular}
	\\
	Notes: redshifts are photometric redshifts from \citet{zou19,zou20}. For ORC4 we have \\ adopted their redshift in preference to the redshift of 0.385 quoted by \citet{norris20}.
\end{table*}

\begin{figure*}
\begin{center}
	\includegraphics[width=0.5\textwidth]{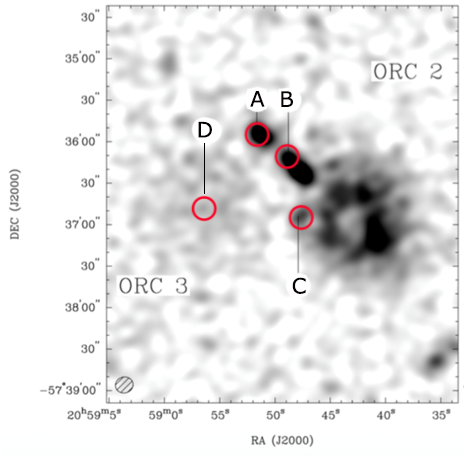}
    \caption{
    ASKAP radio continuum image of ORCs 2 and 3 at 944 MHz, from the EMU Pilot Survey \citep{norris20}, with the  synthesized beam shown in the bottom left corner. ORC2 can be clearly seen on the right hand side, while ORC3 is much fainter, appearing on the left. Radio sources that correspond to putative host galaxies for ORC2 are circled in red and labelled  ``B'', ``C'', and``D''. All these galaxies have redshifts between $z \sim 0.32 - 0.62$. ``A'' is a bright star that is not visible in the radio image.}
    \label{fig:askap_orc3}
\end{center}
\end{figure*}

The remaining two ORCs (ORCs 2 and 3), shown in Figure \ref{fig:askap_orc3}, form a pair whose centres are separated by about 2 arcmin.  Since the other ORCs have a spatial sky density of $\sim$ 1 per 50 sq. deg., the probability of  two ORCs being located within  2 arcmin by chance  is $\sim 10^{-4}$. We therefore assume they are associated, and presumably have a common origin. ORC2 has a near circular, edge-brightened, filled morphology. The diameter is $\sim$ 80 arcsec. Unlike the single ORCs, the diffuse emission is not confined to the circle, but spills outside the ring. ORC3 is much fainter, appearing as a diffuse blob. 

Close to ORC 2/3, there is a 
double-lobed AGN hosted by the galaxy
\mbox{WISEA J205848.80--573612.1} (marked as ``B'' in Figure \ref{fig:askap_orc3}) at a redshift of $z~\sim~0.32$. Coincident with one of the AGN jets is a bright star   \mbox{WISEA~J205851.65--573554.1} (marked as A),  which is a chance association.
 Radio source C is \mbox{WISEA~J205847.91--573653.8}, an edge-on spiral galaxy at a redshift of $z \sim 0.28$. Any of these might be the host galaxy of the ORCs. The ORCs could perhaps be caused by an outflow from the AGN associated with B, or an outflow from the edge-on spiral galaxy C. In the centre of ORC3 is the galaxy  WISEA J205856.25-573644.6 (marked D), at a redshift of $z \sim 0.62$, which might either be a chance background source, or may be an important component of this system. This system will be discussed in detail by Macgregor et al. (in preparation).

The double ORC 2/3
has  properties quite different from the three single ORCs, suggesting that different mechanisms are probably responsible for the two classes.  Unfortunately, incomplete photometry (and hence photometric redshifts) close to ORC2/3 prevent us from including them in the analysis presented here.

In the rest of this paper we focus on the three single sources.

\citet{norris22} argue that the  rings of emission  represents a spherical shell of synchrotron emission surrounding the host galaxy, and  consider two  hypotheses for the origin of this shell:
\begin{itemize}
    \item it is a spherical shock wave  from a cataclysmic event in the host galaxy, such as a merger of two SMBHs, 
    \item it is the termination shock of a starburst wind from past starburst activity in the host galaxy.
\end{itemize}
Other suggested explanations for ORCs include radio jets from an AGN seen end-on (Shabala et al., in preparation)
and the throats of wormholes \citep{kirillov20, kirillov21}.

In this paper we focus on the environment surrounding these host galaxies.

\section{Environment}
\label{environment}

To study the environment of the ORCs, we use the distribution of the photometric redshifts of the nearby galaxies, using the photometric redshift data from the Legacy Survey of the Dark Energy Spectroscopic Instrument (DESI) \citep{zou19,zou20}. These redshifts have a typical quoted standard error of 0.05, and comparisons (Norris, unpublished data) with spectroscopic redshifts confirm these uncertainties are reliable, with very few outliers.

\begin{figure*}
\includegraphics[width=17cm, angle=0]{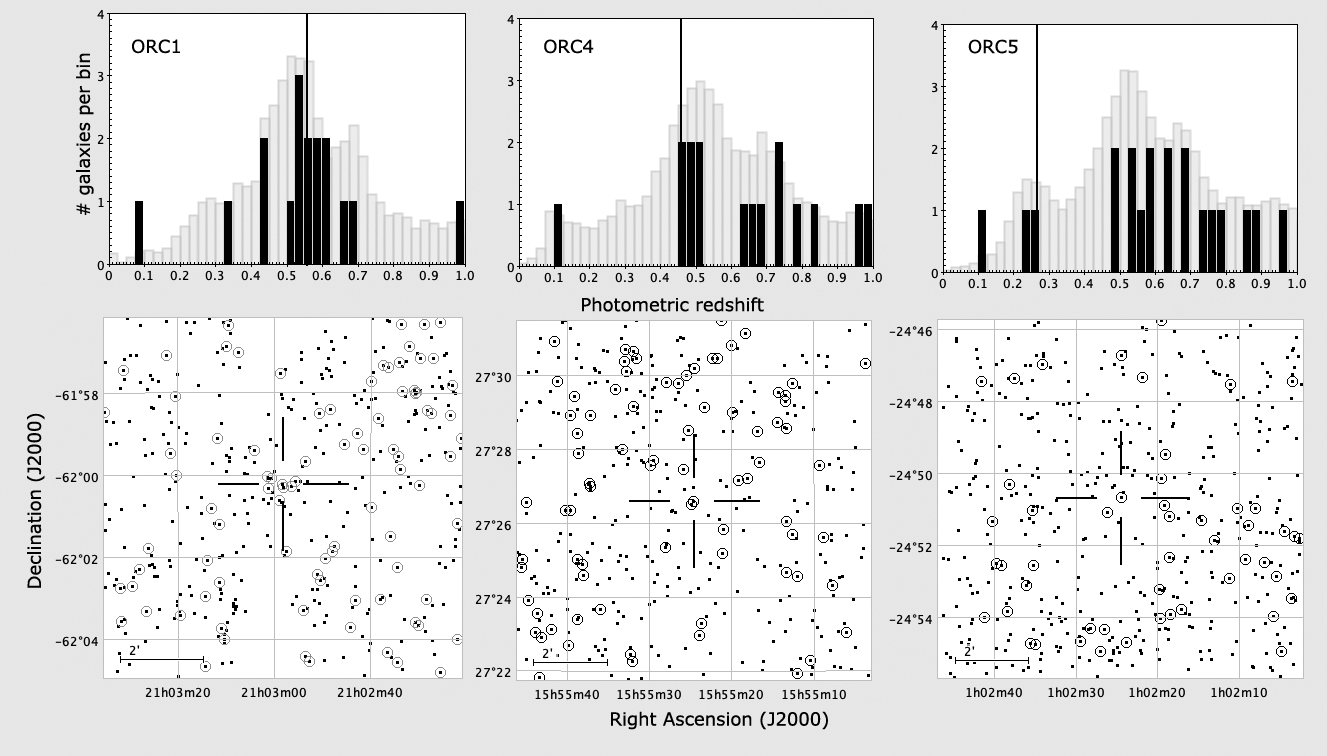}
\caption{(Top) A histogram of the photometric redshifts \citep{zou20} of the galaxies surrounding the central galaxy of each ORC. Dark columns show galaxies within 1 arcmin of the central galaxy, or in the case of ORC1, within 0.5 arcmin of the central galaxy. Grey  columns show the (scaled) distribution of redshifts over a region of a few square degrees surrounding the ORC. The vertical line shows the redshift of the central galaxy. (Bottom) The galaxies surrounding each ORC. Dots show all galaxies with a photometric redshift in \citet{zou20}, and circles show those with a redshift within 0.05 of that of the central galaxy. The cross-hairs show the location of the central galaxy.
}
\label{cluster}
\end{figure*}

In Figure \ref{cluster} we show the distribution of galaxies surrounding each ORC. The top row shows the distribution of redshifts of galaxies within 1 arcmin (or for ORC1, 0.5 arcmin) of the host galaxy, and the bottom row shows the projected spatial distribution of galaxies that have a redshift within $\pm$ 0.05 of that of the host galaxy.

At the left of the diagram, ORC1 is clearly located in an overdensity of galaxies. There are 8 galaxies with $0.5 < z < 0.6$  within 0.5 arcmin of the host galaxy, which gives a spatial density 9 times higher than galaxies in that redshift range in the surrounding area. We chose the radius of 0.5 arcmin to emphasise the high overdensity close to this source.

The individual photometric redshifts of those 8 close galaxies do not have enough precision to accurately locate them in the radial direction (i.e. along the line of sight) relative to the ORC. However, if the radial distribution  is similar to the tangential (i.e. perpendicular to the line of sight) distribution, then most of these galaxies lie within a few 100 kpc of the host galaxy, and several  are presumably physically located within the shell of the ORC. The implications of this are discussed further by  \citet{norris22}.

ORC4 does not exhibit a similar overdensity. However, at the centre of the lower middle plot can be seen a pair of galaxies almost coincident and at a similar redshift   (i.e. the host at 15$^h$ 55$^m$ 24.65$^s$, +27\degr 26'33''.7, z=0.457, and a nearby galaxy at 15$^h$ 55$^m$ 24.85$^s$,    +27\degr26'29''.3, z=0.494). This pair of galaxies has an apparent separation of only 6 arcsec or 35kpc, so they are likely to be an interacting pair. 

ORC5 also does not exhibit an overdensity. However, this too has a companion (01$^h$ 02$^m$ 26.17$^s$,    -24\degr51'05''.1, z=0.246) to the host galaxy (01$^h$ 02$^m$ 24.33$^s$, -24\degr50'39''.5, z=0.270). This companion is visible in Figure \ref{3orcs} as the bright galaxy in the south-east limb of  ORC5. This galaxy has a tangential separation of  36 arcsec or 150kpc from the host, but there is some indication \citep{koribalski21} that the companion is interacting with the ORC.

\section{Discussion}
Most of the models so far proposed for ORCs \citep{norris22} depend critically on the ambient density and magnetic field of the region surrounding the ORC. For example, a magnetic field comparable to the cosmic microwave background equipartition field (e.g. 8$\mu$G for ORC1 \citep{norris22}) is required to generate the observed synchrotron emission. Although an ambient magnetic field will be amplified by the shock that created the ORC, an initial ambient magnetic field is still required to seed that process, and ambient magnetic fields are known to be higher in clusters and over-densities than in a non-cluster environment \citep{norris22}. Thus this model suggests that ORCs should occur preferentially  in overdensity regions.

Of the three ORCs discussed here, one is located in a significant overdensity, and the other two appear to have a nearby, interacting companion. Studies \citep[e.g.][]{yamauchi08} suggest that only a few percent of galaxies have an interacting companion, and so it appears that the host galaxies of ORCs are atypical. However, we are very cautious in making this claim, as (a) our sample of three is very small, and (b) the technique we have used to find companions may probe more deeply than the techniques used in large-scale searches for companions.

Nevertheless, if this apparent atypicality is supported by further work, it may offer a clue as to why ORCs are so rare (about one every 50 square degrees). 

\section{Conclusions}

We have shown that each of the three known single ORCs is either located in an overdensity of nearby galaxies, or it has a close companion with which it is presumably interacting. By comparison, only a few percent of field galaxies have a known companion galaxy. We hesitate to place too much weight on a sample of three objects, but if this result is supported by further studies then it suggests an explanation why only one ORC is found per 50 square degrees of sky.

\vspace{6pt}



\begin{acknowledgements}
We thank B\"{a}rbel Koribalski for comments on an early draft of this paper. This work makes use of data products from the Wide-field Infrared Survey Explorer, which is a joint project of the University of California, Los Angeles, and the Jet Propulsion Laboratory/California Institute of Technology, funded by the National Aeronautics and Space Administration.
It also uses public archival data from the Dark Energy Survey (DES) and we acknowledge the institutions listed on \url{https://www.darkenergysurvey.org/the-des-project/data-access/}.
This research has made extensive use of the ``Aladin sky atlas'' (developed at CDS, Strasbourg Observatory, France \citep{aladin}),  TOPCAT \citep{topcat} and Ned Wright's cosmology calculator \citep{nedwright}.
\end{acknowledgements}

\bibliographystyle{pasa-mnras}
\bibliography{main.bib}

\begin{thebibliography}{}
\makeatletter
\relax
\def\mn@urlcharsother{\let\do\@makeother \do\$\do\&\do\#\do\^\do\_\do\%\do\~}
\definecolor{darkblue}{rgb}{0,0,0.597656}
\def\mndoi{\begingroup\mn@urlcharsother \@ifnextchar [ {\mndoi@} {\mndoi@[]}}
\def\mndoi@[#1]#2{\def\@tempa{#1}\ifx\@tempa\@empty \href
  {http://dx.doi.org/#2} {\textcolor{darkblue}{doi:#2}}\else \href
  {http://dx.doi.org/#2} {\textcolor{darkblue}{#1}}\fi \endgroup}
\def\mn@eprint#1#2{\mn@eprint@#1:#2::\@nil}
\def\mn@eprint@arXiv#1{\href {http://arxiv.org/abs/#1} {{\tt arXiv:#1}}}
\def\mn@eprint@dblp#1{\href {http://dblp.uni-trier.de/rec/bibtex/#1.xml}
  {dblp:#1}}
\def\mn@eprint@#1:#2:#3:#4\@nil{\def\@tempa {#1}\def\@tempb {#2}\def\@tempc
  {#3}\ifx \@tempc \@empty \let \@tempc \@tempb \let \@tempb \@tempa \fi \ifx
  \@tempb \@empty \def\@tempb {arXiv}\fi \@ifundefined
  {mn@eprint@\@tempb}{\@tempb:\@tempc}{\expandafter \expandafter \csname
  mn@eprint@\@tempb\endcsname \expandafter{\@tempc}}}

\bibitem[\protect\citeauthoryear{{Abbott} et~al.,}{{Abbott}
  et~al.}{2018}]{abbott18}
{Abbott} T.~M.~C.,  et~al., 2018, \mndoi [\apjs] {10.3847/1538-4365/aae9f0},
  \href {https://ui.adsabs.harvard.edu/abs/2018ApJS..239...18A} {239, 18}

\bibitem[\protect\citeauthoryear{{Ananthakrishnan} \& {Pramesh
  Rao}}{{Ananthakrishnan} \& {Pramesh Rao}}{2001}]{gmrt}
{Ananthakrishnan} S.,  {Pramesh Rao} A.,  2001, in 2001 Asia-Pacific Radio
  Science Conference AP-RASC '01. p.~237

\bibitem[\protect\citeauthoryear{{Boch} \& {Fernique}}{{Boch} \&
  {Fernique}}{2014}]{aladin}
{Boch} T.,  {Fernique} P.,  2014, in {Manset} N.,  {Forshay} P.,  eds,
  Astronomical Society of the Pacific Conference Series Vol. 485, Astronomical
  Data Analysis Software and Systems XXIII. p.~277

\bibitem[\protect\citeauthoryear{{Hotan} et~al.,}{{Hotan}
  et~al.}{2021}]{hotan21}
{Hotan} A.~W.,  et~al., 2021, \mndoi [\pasa] {10.1017/pasa.2021.1}, \href
  {https://ui.adsabs.harvard.edu/abs/2021PASA...38....9H} {38, e009}

\bibitem[\protect\citeauthoryear{{Kirillov} \& {Savelova}}{{Kirillov} \&
  {Savelova}}{2020}]{kirillov20}
{Kirillov} A.~A.,  {Savelova} E.~P.,  2020, \mndoi [European Physical Journal
  C] {10.1140/epjc/s10052-020-8395-7}, \href
  {https://ui.adsabs.harvard.edu/abs/2020EPJC...80..810K} {80, 810}

\bibitem[\protect\citeauthoryear{{Kirillov}, {Savelova}  \&
  {Vladykina}}{{Kirillov} et~al.}{2021}]{kirillov21}
{Kirillov} A.~A.,  {Savelova} E.~P.,   {Vladykina} P.~O.,  2021, \mndoi
  [Universe] {10.3390/universe7060178}, \href
  {https://ui.adsabs.harvard.edu/abs/2021Univ....7..178K} {7, 178}

\bibitem[\protect\citeauthoryear{{Koribalski}, {Norris}, {Andernach},
  {Rudnick}, {Shabala}, {Filipovi{\'c}}  \& {Lenc}}{{Koribalski}
  et~al.}{2021}]{koribalski21}
{Koribalski} B.~S.,  {Norris} R.~P.,  {Andernach} H.,  {Rudnick} L.,  {Shabala}
  S.,  {Filipovi{\'c}} M.,   {Lenc} E.,  2021, \mndoi [\mnras]
  {10.1093/mnrasl/slab041}, \href
  {https://ui.adsabs.harvard.edu/abs/2021MNRAS.tmpL..42K} {}

\bibitem[\protect\citeauthoryear{{Norris} et~al.,}{{Norris} et~al.}{2011}]{emu}
{Norris} R.~P.,  et~al., 2011, \mndoi [\pasa] {10.1071/AS11021}, \href
  {http://adsabs.harvard.edu/abs/2011PASA...28..215N} {28, 215}

\bibitem[\protect\citeauthoryear{{Norris} et~al.,}{{Norris}
  et~al.}{2021a}]{norris20}
{Norris} R.~P.,  et~al., 2021a, \mndoi [\pasa] {10.1017/pasa.2020.52}, \href
  {https://ui.adsabs.harvard.edu/abs/2021PASA...38....3N} {38, e003}

\bibitem[\protect\citeauthoryear{{Norris} et~al.,}{{Norris}
  et~al.}{2021b}]{norris21}
{Norris} R.~P.,  et~al., 2021b, \mndoi [\pasa] {10.1017/pasa.2021.42}, \href
  {https://ui.adsabs.harvard.edu/abs/2021PASA...38...46N} {38, e046}

\bibitem[\protect\citeauthoryear{{Norris} et~al.,}{{Norris}
  et~al.}{2022}]{norris22}
{Norris} R.~P.,  et~al., 2022, \mnras, p. in preparation

\bibitem[\protect\citeauthoryear{{Taylor}}{{Taylor}}{2005}]{topcat}
{Taylor} M.~B.,  2005, in {Shopbell} P.,  {Britton} M.,   {Ebert} R.,  eds,
  Astronomical Society of the Pacific Conference Series Vol. 347, Astronomical
  Data Analysis Software and Systems XIV. p.~29

\bibitem[\protect\citeauthoryear{{Wright}}{{Wright}}{2006}]{nedwright}
{Wright} E.~L.,  2006, \mndoi [\pasp] {10.1086/510102}, \href
  {https://ui.adsabs.harvard.edu/abs/2006PASP..118.1711W} {118, 1711}

\bibitem[\protect\citeauthoryear{Yamauchi, Yagi  \& Goto}{Yamauchi
  et~al.}{2008}]{yamauchi08}
Yamauchi C.,  Yagi M.,   Goto T.,  2008, \mndoi [Monthly Notices of the Royal
  Astronomical Society] {10.1111/j.1365-2966.2008.13756.x}, 390, 383

\bibitem[\protect\citeauthoryear{{Zou}, {Gao}, {Zhou}  \& {Kong}}{{Zou}
  et~al.}{2019}]{zou19}
{Zou} H.,  {Gao} J.,  {Zhou} X.,   {Kong} X.,  2019, \mndoi [\apjs]
  {10.3847/1538-4365/ab1847}, \href
  {https://ui.adsabs.harvard.edu/abs/2019ApJS..242....8Z} {242, 8}

\bibitem[\protect\citeauthoryear{{Zou}, {Gao}, {Zhou}  \& {Kong}}{{Zou}
  et~al.}{2020}]{zou20}
{Zou} H.,  {Gao} J.,  {Zhou} X.,   {Kong} X.,  2020, VizieR Online Data
  Catalog, \href {https://ui.adsabs.harvard.edu/abs/2020yCat..22420008Z} {p.
  J/ApJS/242/8}

\makeatother
\end{thebibliography}

\end{document}